\documentclass[journal]{IEEEtran}
\usepackage{amsmath,amsfonts,amssymb}
\usepackage{algorithmic}
\usepackage{algorithm}
\usepackage{array}
\usepackage{textcomp}
\usepackage{stfloats}
\usepackage{url}
\usepackage{verbatim}
\usepackage{graphicx}
\usepackage{subfigure}
\usepackage{cite}
\usepackage{multirow}
\usepackage{pifont}
\usepackage{float}
\usepackage[]{colortbl}
\usepackage[table]{xcolor}
\usepackage{subfigure, multirow, url, amsfonts, amssymb, pifont, booktabs, siunitx, makecell, hyperref}
\usepackage{xcolor}
\newcommand{\rev}[1]{#1}
\newcommand{\rrev}[1]{#1}
\begin{document}

\title{UL-UNAS: Ultra-Lightweight U-Nets for Real-Time Speech Enhancement via Network Architecture Search}
\author{Xiaobin Rong, Leyan Yang, Dahan Wang, Yuxiang Hu, Changbao Zhu, Kai Chen, Jing Lu,~\IEEEmembership{Senior Member,~IEEE}

\thanks{This work was supported by the National Natural Science Foundation of China (Grant No. 12274221), Yangtze River Delta Science and Technology Innovation Community Joint Research Project (Grant No. 2024CSJGG1103), and the AI \& AI for Science Project of Nanjing University. (Corresponding author: Jing Lu.)

The authors are with the Key Laboratory of Modern Acoustics, Institute of Acoustics, Nanjing University, Nanjing 210093, China, and also with the NJU-Horizon Intelligent Audio Lab, Horizon Robotics, Beijing 100094, China (e-mail: xiaobin.rong@smail.nju.edu.cn; leyan.yang@smail.nju.edu.cn; dahan.wang@smail.nju.edu.cn; yuxiang.hu@gua.com; changbao.zhu@gua.com; chenkai@nju.edu.cn; lujing@nju.edu.cn).}}

\maketitle

\begin{abstract}
Lightweight models are essential for real-time speech enhancement applications. In recent years, there has been a growing trend toward developing increasingly compact models for speech enhancement. In this paper, we propose an Ultra-Lightweight U-Net optimized by Network Architecture Search (UL-UNAS), which is suitable for implementation in low-footprint devices. Firstly, we explore the application of various efficient convolutional blocks within the U-Net framework to identify the most promising candidates. Secondly, we introduce two boosting components to enhance the capacity of these convolutional blocks: a novel activation function named affine PReLU and a causal time-frequency attention module. Furthermore, we leverage neural architecture search to discover an optimal architecture within our carefully designed search space. By integrating the above strategies, UL-UNAS not only significantly outperforms the latest ultra-lightweight models with the same or lower computational complexity, but also delivers competitive performance compared to recent baseline models that require substantially higher computational resources. \rrev{Source code and audio demos are available at \url{https://github.com/Xiaobin-Rong/ul-unas}.}
\end{abstract}


\section{Introduction}
\IEEEPARstart{S}{peech} enhancement (SE) aims to recover clean speech from the noisy mixture, thereby improving speech quality and intelligibility. In recent years, significant breakthroughs in SE have been achieved due to the rapid evolution of deep neural networks (DNNs). In general, DNN-based SE algorithms can be categorized into the time-frequency (T-F) domain approaches \cite{CRN, DCCRN, DPCRN, BSRNN, DeepFilterNetV2, TF-GridNet} and the time-domain approaches \cite{TCNN, ConvTasNet, DPRNN, TSTNN}. For the T-F domain approaches, the short-time Fourier transform (STFT) representation of noisy speech is fed into a DNN, which outputs an enhanced speech spectrogram through techniques such as masking \cite{CRN, DCCRN, DPCRN, BSRNN}, deep filtering \cite{DeepFilterNetV2, Deep_Filtering}, or mapping \cite{TF-GridNet}. Phase estimation is a tricky problem in these cases, as the phase spectrum usually presents an irregular structure due to the wrapping effect. Earlier T-F domain-based works only \rev{focused} on the recovery of spectral magnitude by using the ideal ratio mask (IRM) \cite{IRM}, leaving the phase unaltered. In contrast, phase-sensitive mask \cite{Phase-sensitive} takes into account the cosine of the phase difference between clean and noisy speech when estimating IRM and shows improved results. Complex ideal ratio mask (cIRM) \cite{CRM} applies a complex instead of a real gain to perform implicit phase correction, leading to better perceptual quality but negligible improvements in objective intelligibility. Directly predicting the real and imaginary parts of the clean spectrum can also implicitly recover the phase information, and usually has a higher upper bound of performance than that of masking methods \cite{Complex_spectral_mapping, TF-GridNet}.

Compared to the T-F domain approaches, time domain approaches circumvent the challenging phase estimation problem by conducting feature encoding and decoding in a trainable manner. They can achieve very low algorithm latency due to the learnable signal analysis and synthesis on very short windows. However, they sacrifice the advantages of STFT, which provides a feasible and robust feature representation where speech and noise components exhibit distinct structural characteristics in the spectral domain. Consequently, the performance of time domain methods is not among the top tier in SE tasks \cite{PHASEN}.

Although DNN-based approaches have achieved overwhelming performance over traditional SE algorithms, their improved performance is often accompanied by large computation overhead. Most state-of-the-art (SOTA) SE models require substantial computational resources, making them infeasible for deployment on edge devices for practical applications. Therefore, exploring lightweight SE neural networks with performance comparable to the SOTA models is increasingly becoming a research hotspot. One straightforward solution is model compression, including pruning \cite{Pruning3}, quantization \cite{Towrards_SE_Compression}, and inference skipping \cite{Skip-DPCRN}. TinyLSTMs \cite{TinyLSTMs} applies these three techniques to a simple long short-term memory (LSTM) SE network, achieving a 2.9$\times$ reduction in operations with minor performance degradation. 

Another commonly adopted approach is to hand-craft efficient architectures. TRU-Net \cite{TRUNet} replaces the standard convolution with depthwise separable convolution and employs a more powerful dual-path recurrent neural network (RNN) structure. In \cite{EfficientSE}, it is demonstrated that parallel RNN grouping and convolutional additive skip connections are highly effective in the design of compact architectures, enabling the corresponding model to stand out within the family of manually crafted architectures. Furthermore, combining the advantages of traditional signal processing and DNNs has demonstrated remarkable potential. RNNoise \cite{RNNoise} and PercepNet \cite{PercepNet} first perform a coarse enhancement on a low-resolution spectral envelope and then apply finer noise suppression between pitch harmonics using a pitch comb filter. Similarly, DeepFilterNet \cite{DeepFilterNet, DeepFilterNetV2} performs enhancement on compact ERB features while reconstructing periodic components through deep filtering utilizing coefficients predicted from both ERB and the original STFT features. DPCRN-CF \cite{DPCRN-CF} utilizes a DNN-based pitch estimator and a learnable comb filter to achieve superior harmonic enhancement. Despite the significant reduction in computational overhead achieved by these works, they are still too large for practical deployment on edge devices with low power consumption requirements, except for RNNoise, which is compact enough but suffers from limited performance.

On the other hand, neural architecture search (NAS) is an approach to automating the model design process. NAS aims to find the optimal architecture based on a specific performance estimation strategy within the defined search space, using an appropriate search strategy \cite{NAS_survey}. NASNet \cite{NASNet} employs a controller network to sample architecture candidates from a certain search space. Subsequently, each candidate is trained and evaluated to obtain accuracy as a reward. Eventually, the controller is updated through a reinforcement learning (RL) algorithm. MnasNet \cite{MnasNet} goes further by introducing a novel factorized search space, along with taking latency into account when computing reward, thereby facilitating the design of a platform-aware low-latency network. EfficientNet \cite{EfficientNet} employs the same NAS strategy as MnasNet with the exception that it focuses on optimizing floating-point operations (FLOPs) instead of latency, thereby preventing itself from being restricted to any particular hardware device. 

The number of parameters and computational complexity, represented by floating-point operations per second (FLOPS) or multiply-accumulate operations per second (MACS), serve as the principal guidelines for researchers in the design of lightweight models. The number of parameters determines the memory footprint of models and holds particular significance when the objective is to deploy models on memory-constrained devices. However, in most cases, it is latency, rather than memory, that acts as the bottleneck of edge devices. In such cases, computational complexity can offer a more pertinent and useful metric than the number of parameters. When comparing different architectures, the number of parameters does not offer particularly informative insights, as different architectures typically exhibit varying levels of parameter efficiency \cite{SE_scaling}. For instance, a fully connected network can have massive parameters yet still manage to achieve a relatively fast inference speed. In contrast, a transformer network may have fewer parameters but run much slower due to the computationally expensive attention operations. 
For models of the same architecture type, increasing the number of parameters can lead to improved performance. ParameterNet \cite{ParameterNet} demonstrates this in the context of large-scale visual pretraining. By leveraging dynamic convolution to add parameters with only a marginal increase in FLOPS, it achieves superior results.

In this paper, we focus on optimizing the ultra-lightweight SE model with computational complexity as low as around 30M MACS while imposing no strict constraints on the number of parameters. We adopt the T-F domain approach due to its performance advantage under low computational resource constraints. Building on our previously proposed ultra-lightweight SE model, GTCRN \cite{GTCRN}, we further explore convolutional encoder-decoder (CED) designs to improve both capacity and efficiency of the U-Net architecture. In addition, we apply NAS to refine its structure and identify an optimal configuration. Our contributions are summarized as follows:
\begin{itemize}
    \item {We systematically investigate and evaluate diverse efficient convolutional modules within an ultra-lightweight SE model and identify the most promising designs. This work provides insights for balancing performance and computational efficiency in the development of ultra-lightweight SE models.}
    \item {We develop two strategies to augment the model capacity: a novel nonlinear activation termed affine PReLU (APReLU), and an improved lightweight attention mechanism named causal time-frequency attention (cTFA). Comprehensive ablation studies demonstrate the effectiveness of both proposed methods.}
    \item {We introduce the NAS technique to find an outstanding architecture within our elaborately designed search space. By formulating a computational complexity-aware optimization objective, we ensure that the discovered architecture achieves superior performance while maintaining computational efficiency.}
    \item {Our proposed UL-UNAS, with lower computational complexity, outperforms existing SOTA ultra-lightweight SE models. With only \rev{35M} MACS and a compact size of \rev{171k} parameters, UL-UNAS achieves a PESQ score of 3.09 on the VCTK-DEMAND dataset. This advancement paves the way for more efficient and effective ultra-lightweight designs in various application scenarios.}
\end{itemize}

\section{Related Work}
In this section, we briefly review previous works related to our proposed method and highlight the key distinctions.

\subsection{Ultra-lightweight speech enhancement models}
Over the past few years, the concept of ultra-lightweight SE neural networks has gradually emerged, aiming to be compatible with edge devices with extremely limited computing resources. 
\rev{Several seminal ultra-lightweight SE models include GTCRN \cite{GTCRN} (34M MACS), FSPEN \cite{FSPEN} (89M MACS), and LiSenNet \cite{LiSenNet} (56M MACS). Among these, LiSenNet achieves the highest performance while maintaining a relatively low computational cost.}

All the aforementioned models are built upon the U-Net backbone with a dual-path bottleneck enhancer, thanks to its computational efficiency. Various modifications have been introduced into this base architecture to further enhance performance. The modifications are centered around two key components: \textit{the CED structure} and \textit{the bottleneck enhancer}. Regarding the CED component, GTCRN incorporates grouped convolution along with lightweight boosting modules, including sub-band feature extraction and lightweight attention, to strengthen the basic building blocks. FSPEN utilizes a dual-CED architecture comprising sub-band and full-band processing for better speech feature extraction. LiSenNet introduces novel sub-band downsampling and upsampling blocks specifically designed for band-aware feature capture. As for the bottleneck enhancer, GTCRN adopts a parallel RNN grouping strategy to reduce computational burden. FSPEN develops a path extension method to improve inter-frame modeling. LiSenNet introduces a gate mechanism at the end of the dual-path architecture to augment model capacity.
\rev{Given that GTCRN exhibits the lowest computational cost and best matches our target (around 30M MACS), we select it as the backbone} and focus on exploring more efficient designs for the building blocks within the CED structure.

\subsection{Efficient designs of convolutional blocks}
\label{sec:efficient_convolution_design}
Numerous innovative concepts have been introduced to design efficient convolutional networks to improve their efficacy. According to \cite{StarNet}, these insights can be divided into several categories. Among them, depthwise separable convolution \cite{MobileNet, MobilenetV2, MobileNetV3} decomposes a standard convolution into a depthwise and a pointwise convolution. Feature shuffling \cite{ShuffleNet, ShuffleNetV2} simplifies pointwise convolution by group convolution and channel shuffle. Feature re-use \cite{GhostNet} generates additional feature maps through inexpensive linear transformation, enabling more efficient information exploitation. Reparameterization \cite{MobileOne} introduces multiple parallel branches in training to improve performance, which are merged into one branch before inference. Star operation \cite{StarNet} augments the representative ability of a convolutional block by mapping inputs into a high-dimensional non-linear feature space. 
In this paper, we explore the potential applications of these designs in ultra-lightweight SE neural networks, as detailed in Sec.~\ref{sec:base_blocks}.

\subsection{Nonlinear activation functions}
The design of nonlinear activation functions has been an active area of research in neural networks, as it directly impacts the capabilities of deep learning models. The early proposed smooth nonlinear activation functions, such as Sigmoid and Tanh, suffer from the problem of vanishing gradients, which constrains the performance of DNNs. The rectified linear unit (ReLU) \cite{ReLU}, by presenting a non-saturating nonlinearity, alleviates this issue and has emerged as a preferred option in modern networks. To tackle the ``dying ReLU'' problem, \rev{in which standard ReLU units may become inactive when their inputs fall entirely within the negative region}, some variants have been proposed, such as leaky ReLU \cite{LeakyReLU} and parametric ReLU (PReLU) \cite{PReLU}, by permitting a small, non-zero gradient when the input is negative. Utilizing a weighted combination of base activation functions has also been explored. Adaptive piecewise linear activation unit \cite{APL_act} computes the activation output through a weighted sum of multiple ReLU units with learnable weights and bias. VanillaNet \cite{Vanillanet} introduces a series-informed activation function to enhance the nonlinearity of a neural network. It achieves this by concurrently stacking the activation functions, which can be regarded as a series in a mathematical sense. To further enrich the approximation ability of the series, the spatial neighbors of a specific element of the inputs are taken into account to produce outputs. However, this causes it to function in a manner analogous to a depthwise convolution operation, leading to an increase in computational complexity. In contrast, our proposed APReLU (as described in Sec.~\ref{sec:aprelu}) integrates a learnable affine transformation with PReLU, creating a more powerful activation function with negligible computational overhead.

\subsection{Efficient attention modules}
Attention modules are found to be effective in boosting the performance of neural networks. They typically operate by aggregating information across selected dimensions to produce a compact, context-rich representation, enabling the model to focus on salient features along the target dimension. This process typically incurs low computational cost due to dimensionality reduction. SENet \cite{SENet} \rev{aims to boost the representational capacity of networks by explicitly modeling interdependencies between channels, introducing} a novel channel-attention block that adaptively recalibrates channel-wise feature responses based on input features pooled over spatial dimensions. CBAM \cite{CBAM} \rev{extends this idea by also capturing spatial interdependencies}, incorporating a cascaded spatial attention module based on input features pooled over channel dimensions. TFA \cite{TFA_taslp} introduces a similar attention mechanism \rev{for speech enhancement by modeling time and frequency interdependencies in parallel}, achieving a T-F distribution modeling by two parallel attention branches operating on features pooled over the frequency and time dimensions, respectively. However, the average pooling along the time axis within the frequency attention branch results in non-causality, thus hindering its application in real-time SE models. Although the TRA introduced in GTCRN overcomes the non-causality issue by discarding the frequency attention branch, it sacrifices the potential benefits of frequency-dependent attention. In contrast, our proposed cTFA (as described in Sec.~\ref{sec:ctfa}) preserves this frequency-dependent branch while \rev{offering an alternative, lightweight mechanism for modeling frequency dependencies without relying on time-axis pooling.}

\subsection{Neural architecture search for speech enhancement}
While NAS methods have outperformed manually designed architectures on many tasks such as image classification, object detection, and semantic segmentation, there are few works on applying NAS in SE. NAS-TasNet \cite{NAS-TasNet} investigates applying both gradient descent and RL-based NAS methods on a classic speech separation model, Conv-TasNet \cite{ConvTasNet}, and achieves higher performance with fewer parameters. However, the resulting model remains too large to be considered lightweight. In \cite{QA-NAS_SE}, a quantization-aware NAS method is used to design an ultra-low memory SE system with just 9.2k parameters. However, the search space is defined by a trivial U-Net architecture, which limits its performance upper bound and hinders its potential to achieve more advanced capabilities. In this paper, we apply NAS to discover the optimal configuration for constructing an ultra-lightweight U-Net, where the search space is defined by our elaborately designed, high-performance blocks.

\section{Methodology}

\begin{figure}[!t]
    \centering
    \includegraphics[width=\linewidth]{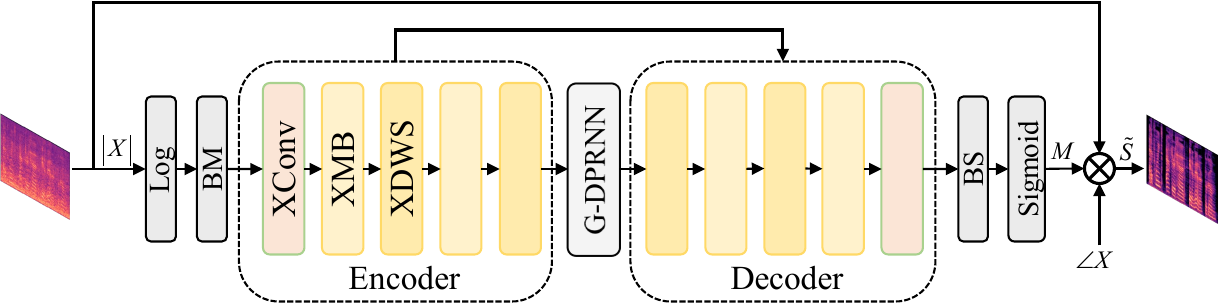}
    \caption{Overall architecture of UL-UNAS.}
    \label{fig:ulunas}
\end{figure}

\subsection{Problem formulation}
We employ the T-F domain approach to build an ultra-lightweight SE model. The noisy speech can be expressed in the T-F domain as:
\begin{equation}
    \label{eq:1}
    X(t,f)=S(t,f)+N(t,f),
\end{equation}
where $X(t,f)$, $S(t,f)$, and $N(t,f)$ denote the complex spectrograms of the noisy speech, the clean speech, and the noise, respectively, with time index $t$ and frequency index $f$. We employ the masking method to obtain the enhanced speech, as it has been shown to perform better than mapping when the model size is limited. Although estimating cIRM is widely adopted in many SE models to recover the phase, we have observed that it is prone to yielding an estimated mask with nearly zero imaginary parts, meaning that it is downgraded to IRM. This is in alignment with the phenomena elucidated in \cite{PHASEN}. We attribute this to the insufficient capacity of the \rev{ultra-lightweight models}. Therefore, we resort to a straightforward yet effective IRM-based masking method to recover the clean spectrogram from the noisy one, which can be expressed as:
\begin{equation}
    \label{eq:2}
    \tilde{S}(t,f)=X(t,f)\otimes M(t,f),
\end{equation}
where $\otimes$ denotes element-wise multiplication and $\tilde{S}(t,f)$ represents the enhanced spectrogram.

\subsection{Model \rev{Overview}}
We adopt the GTCRN framework \rev{as the model backbone}, as illustrated in Fig.~\ref{fig:ulunas}. We retain the band merging (BM), band splitting (BS), and grouped dual-path RNN (G-DPRNN) modules, while replacing the original convolutional blocks with \rev{our elaborately designed} blocks\rev{: XConv, XDWS, and XMB blocks, which are described in the following sections}. For blocks in the decoder, convolutional layers are replaced with transposed variants accordingly. \rev{The number of blocks, their arrangement, and the associated hyperparameter configurations are all determined via NAS. The model takes the noisy log-power spectrogram as input and predicts a magnitude mask, constrained within the range of (0,1), to recover the enhanced magnitude. The enhanced spectrogram is then obtained by combining this magnitude with the original noisy phase.}

\subsection{\rev{Model Components}}
\rev{In this section, we first investigate the most efficient convolutional block designs for SE, as surveyed in Sec.~\ref{sec:efficient_convolution_design}. We then introduce two novel components proposed in this work: APReLU and cTFA, which are designed to augment model representational capacity. The final encoder and decoder architectures are constructed by integrating these components into the selected base blocks.}

\begin{figure}[!t]
    \centering
    \includegraphics[width=\linewidth]{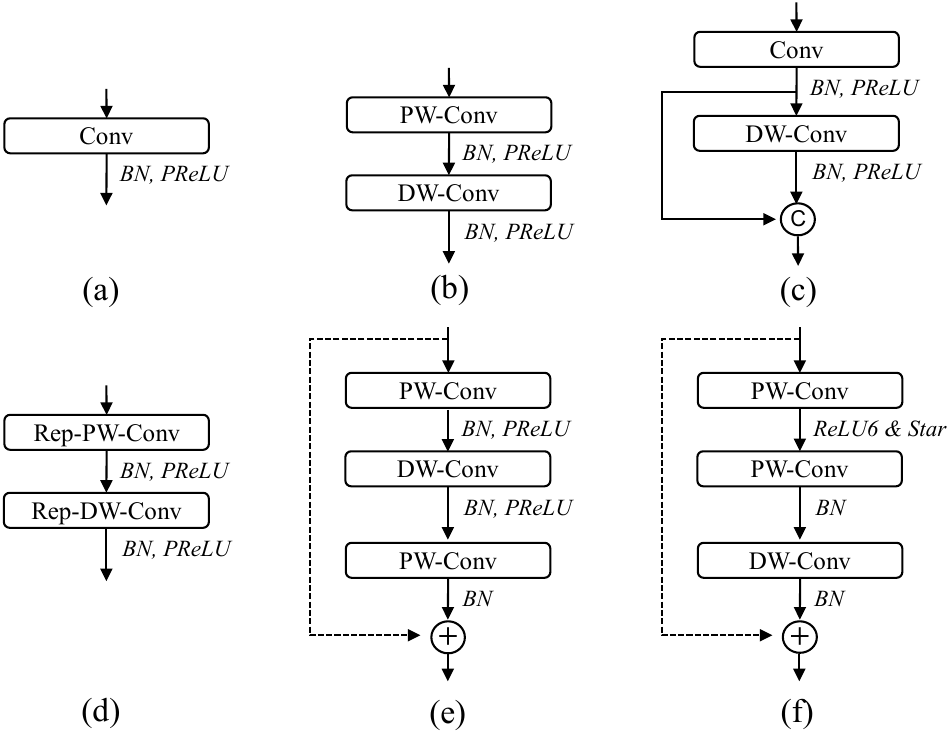}
    \caption{Efficient convolutional blocks modified for ultra-lightweight SE. (a) Conv block. (b) DWS block. (c) Ghost block. (d) Rep block. (e) MB block. (f) Star block. PW and DW refer to pointwise and depthwise, respectively, and Rep represents reparameterizable, with details omitted in this figure.}
    \label{fig:base_blocks}
\end{figure}

\begin{figure}[!t]
    \centering
    \includegraphics[width=\linewidth]{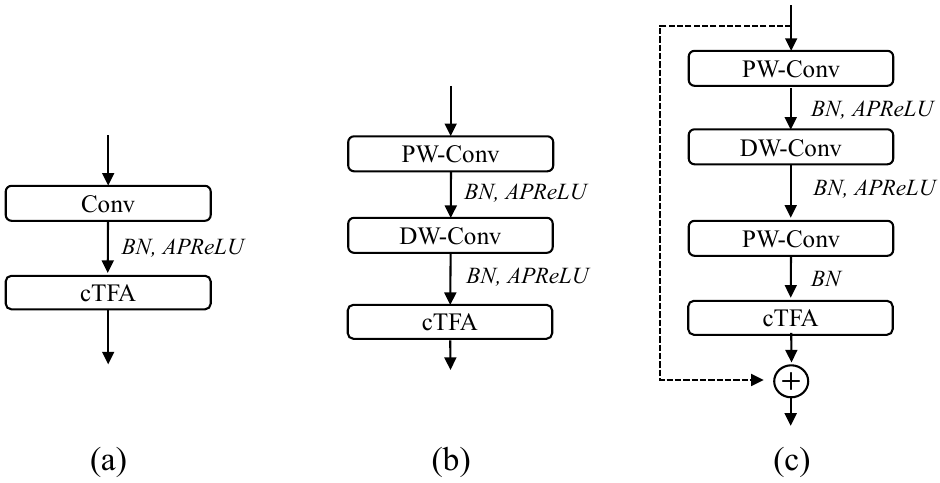}
    \caption{Extended convolutional blocks integrated with APReLU and cTFA. (a) XConv block. (b) XDWS block. (c) XMB block.}
    \label{fig:xblocks}
\end{figure}

\subsubsection{\rev{Base} blocks}
\label{sec:base_blocks}
We explore various efficient convolutional block designs \rev{as follows}.
\begin{itemize}
    \item \textit{Conv block}: A standard convolutional layer followed by a batch normalization (BN) and a PReLU activation;
    \item \textit{DWS block}: The depthwise separable block from MobileNet \cite{MobileNet}, which decomposes a standard convolution to a depthwise and a pointwise convolution.
    \item \textit{Ghost block}: The lightweight block introduced in GhostNet \cite{GhostNet}, which uses a cost-effective depthwise convolution to create ghost features;
    \item \textit{Rep block}: The efficient block introduced in MobileOne \cite{MobileOne}, which applies reparameterization techniques on the DWS block to improve performance without extra computation during inference;
    \item \textit{MB block}: The inverted residual block from MobileNetV2 \cite{MobilenetV2}, incorporating an additional pointwise convolution and residual connection based on the DWS block;
    \item \textit{Star block}: The optimal variant of star blocks explored in StarNet \cite{StarNet}, leveraging the star operation for implicit high-dimensional representative capacity.
\end{itemize}
We introduce several modifications to the original block designs. For the DWS and Rep blocks, we reposition the pointwise convolution before the depthwise convolution, allowing the input to be mapped to a higher-dimensional space at the initial stage. For the Star block, we remove the first depthwise convolution to maintain the same number of layers as in the MB block. In both the MB and Star blocks, an additional BN is added at the end empirically. Besides, the residual connection is applied only when the output feature shape matches the input feature shape. The architectural details of these block designs are illustrated in Fig.~\ref{fig:base_blocks}.

We classify these blocks into three categories based on the number of convolutional layers inside. Within each category, the most promising block is selected for augmentation with our proposed APReLU and cTFA modules, and subsequently serves as a candidate for NAS. \rev{Based on our ablation studies in Sec.~\ref{sec:effects_base_blocks}, the Conv, DWS, and MB blocks were selected, from which we constructed the XConv, XDWS, and XMB blocks, respectively, as illustrated in Fig.~\ref{fig:xblocks}.}

\begin{figure}
    \centering
    \subfigbottomskip=2pt
    \subfigcapskip=-5pt
    \subfigure[]{
        \includegraphics[width=0.45\linewidth]{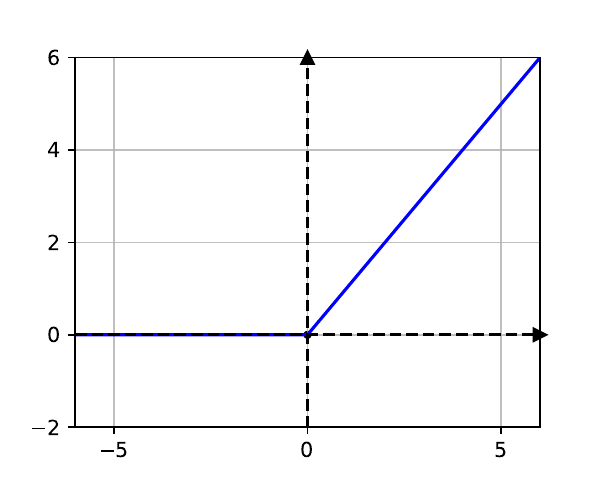}
    }
    \subfigure[]{
        \includegraphics[width=0.45\linewidth]{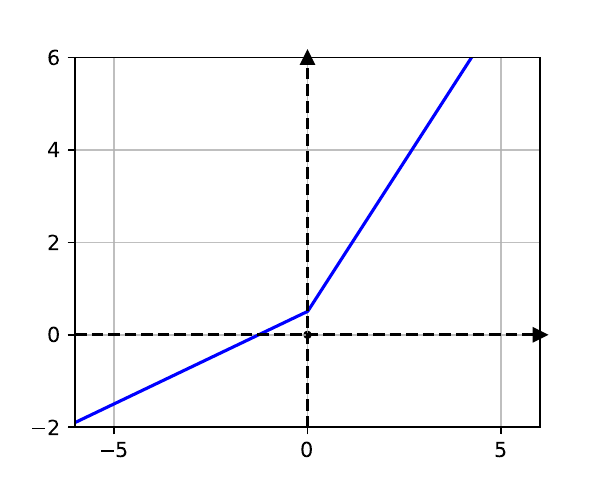}
    }\vspace{3mm}
    \subfigure[]{
        \includegraphics[width=0.45\linewidth]{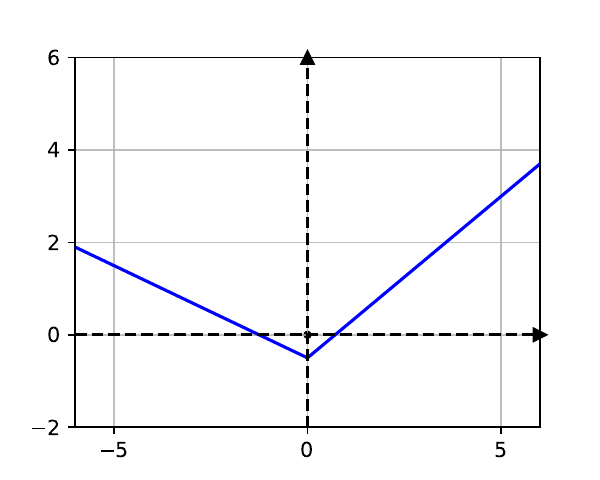}
    }
    \subfigure[]{
        \includegraphics[width=0.45\linewidth]{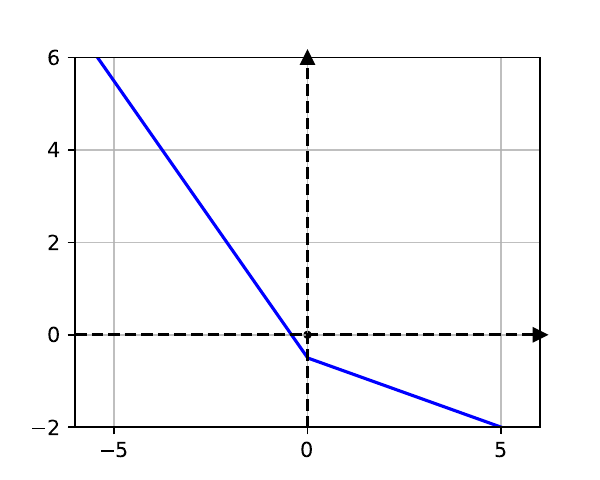}
    }
    \caption{Examples of the proposed APReLU. (a) $\gamma=\beta=\alpha=0$. (b) $\gamma=0.3$, $\beta=0.5$, $\alpha=0.1$. (c) $\gamma=-0.3$, $\beta=-0.5$, $\alpha=-0.1$. (d) $\gamma=-1.3$, $\beta=-0.5$, $\alpha=0.1$.}
    \label{fig:aprelu}
\end{figure}

\subsubsection{Affine PReLU}
\label{sec:aprelu}
\rev{As speech characteristics and noise distributions vary substantially across frequency bands, we argue that a frequency-dependent activation function is more appropriate than conventional frequency-agnostic alternatives for speech enhancement. To this end, we introduce} the APReLU activation function, {which combines a frequency-dependent} affine transformation \rev{with} a PReLU activation, formulated as:
\begin{align}
    h(x)&=\gamma x+\beta+\text{PReLU}(x) \\
        &=\gamma x+\beta+\max(0,x)+\alpha \cdot \min(0,x),
    \label{eq:3-4}
\end{align}
where $x\in \mathbb{R}^{C\times T \times F}$ represents the input feature, with $C,T,F$ denoting the channel, time, and frequency axis lengths, respectively. $\gamma,\beta \in \mathbb{R}^{C\times F}$ is the frequency-dependent scaling factor and bias of the affine transformation, and $\alpha \in \mathbb{R}^{C}$ is the slope-controlling parameter in PReLU. The $\gamma$ coefficient is initialized to all ones and the $\beta$ coefficient to zeros, so that the affine transformation starts as an identity mapping, \rev{which allows the module to start from a residual connection and gradually learn adaptations. The $\alpha$ coefficient is initialized to 0.25 following the commonly used default in PyTorch, providing a stable and widely adopted starting point.} All three coefficients are learnable.

Example APReLU functions are illustrated in Fig.~\ref{fig:aprelu}. As compared to conventional PReLU, our proposed APReLU supports a broader range of activation modes. Notably, the \rev{frequency-dependent} scaling factors and biases \rev{enable the activation to recalibrate the energy across different frequency bands, thereby offering the potential to more effectively enhance speech components while suppressing noise.}

\begin{figure}[!t]
    \centering
    \includegraphics[width=1\linewidth]{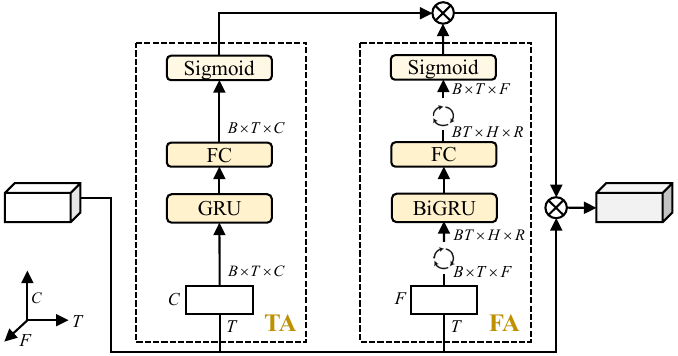}
    \caption{Detailed architecture of the proposed cTFA module.}
    \label{fig:ctfa}
\end{figure}

\subsubsection{Causal time-frequency attention}
\label{sec:ctfa}
TFA \cite{TFA_taslp} is a lightweight attention module for salient T-F speech distribution modeling using two parallel branches, namely time attention (TA) and frequency attention (FA). Concerning an intermediate T-F representation, time-frame-wise and frequency-wise statistical features are respectively generated through average pooling over the frequency and time dimensions. Subsequently, stacked one-dimensional (1D) convolutional layers are employed to model the distribution and produce the attention map for each statistical feature. A final two-dimensional (2D) T-F attention map is obtained via a tensor multiplication operation. Although it demonstrates effectiveness in being flexibly integrated with existing backbone networks to improve performance, the average pooling along the time dimension gives rise to non-causality, thereby hampering its application in real-time SE models. In TRA \cite{GTCRN}, the non-causal branch is discarded so that the final attention map is frequency-independent, which unfortunately leads to a certain degree of performance deterioration. \rev{However, we argue that the essential role of FA lies in modeling frequency dependencies, and this goal does not inherently require time-pooling. Accordingly, our cTFA module specifically replaces the time-pooling-based FA branch with a channel-pooling-based one. This design provides an effective alternative for frequency modeling while inherently ensuring causality.}

Our proposed cTFA module, \rev{built upon TRA, is designed to further capture frequency dependencies and generate a finer-grained attention map. To circumvent the non-causality issue in TFA, we avoid any time-pooling operations and instead introduce a lightweight frequency-attention mechanism based on channel-pooling features.} As shown in Fig.~\ref{fig:ctfa}, \rev{cTFA} consists of two parallel branches: \rev{TA} and \rev{FA}. The \rev{TA} branch is the same as the TRA module, which consists of a GRU layer, a fully connected (FC) layer, and a Sigmoid activation. \rev{The FA branch adopts an analogous architecture but replaces the GRU with a bidirectional GRU (BiGRU) to achieve stronger spectral modeling capability. To balance representational capacity and computational cost, we introduce an efficient \textit{frequency folding} operation. This process transforms the channel-pooled feature map (of size $B \times T \times F$) into a compact representation (of size $BT \times H \times R$) by reshaping the frequency dimension. Specifically, the $F$ frequency bins are consolidated into $H$ frequency groups, each encompassing a local spectral context of size $R$ (fixed to 4 in this work, with $F=H \cdot R$). When $F$ is not divisible by $R$, zero-padding is applied along the frequency dimension.
This folding operation allows the subsequent BiGRU to model long-range dependencies along the $H$ axis and local spectral context along the $R$ axis more effectively and efficiently.}

Given $V\in \mathbb{R}^{B \times C \times T \times F}$ as the input feature, the \rev{time} attention map $A_T \in \mathbb{R}^{B\times C\times T}$ is first computed as follows:
\begin{align}
    Z_T &= \frac{1}{F} \sum_{f=1}^{F} V^2(b, c,t,f), \\
    A_{T} &= \rev{\text{TA}}(Z_T).
    \label{eq:5-6}
\end{align}
On the other hand, the \rev{frequency} attention map $A_F \in \mathbb{R}^{B\times T\times F}$ is computed as follows:
\begin{align}
    Z_F &= \frac{1}{C} \sum_{c=1}^{C} V^2(b, c,t,f), \\
    A_{F} &= \rev{\text{FA}}(Z_F).
    \label{eq:7-8}
\end{align}
The final multi-dimensional attention map $A_{TF} \in \mathbb{R}^{B\times C \times T \times F}$ is computed as:
\begin{equation}
    A_{TF}(b, c,t,f) = A_{T}(b, c,t) \cdot A_{F}(b, t,f).
    \label{eq:9}
\end{equation}
The final output is given as:
\begin{equation}
    \tilde{V}=V\otimes A_{TF},
    \label{eq:10}
\end{equation}
where $\otimes$ denotes the element-wise multiplication operation.

\subsection{Neural architecture search}
\begin{figure}
    \centering
    \includegraphics[width=0.7\linewidth]{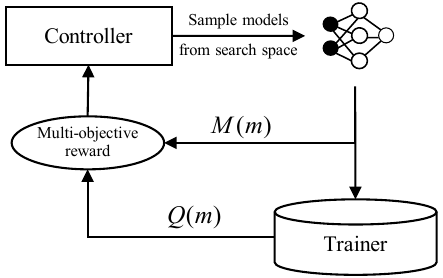}
    \caption{Diagram of MACS-aware neural architecture search.}
    \label{fig:nas}
\end{figure}

We optimize the network configuration within a search space composed of the aforementioned extended blocks using NAS. To improve search efficiency, we design a relatively small search space with respect to the encoder. Once the encoder is determined, the entire network can be ascertained. The encoder is partitioned into a sequence of our pre-defined extended blocks, each spanning a per-block sub-search space. Specifically, the sub-search space for block $i$ comprises 5 configurable nodes, each offering multiple options:
\begin{itemize}
    \item Block type ($T_i)$: XConv, XDWS, or XMB
    \item Stride size ($S_i$): 1 or 2
    \item Group number ($G_i$): 1 or 2
    \item Channel number ($C_i$): 12, 16, 20, 24, 28, 32, or 36
    \item Kernel size ($K_i$): $1\times5$, $1\times7$, $2\times5$, or $3\times3$
\end{itemize}
The stride size of a block refers to the stride along the frequency dimension in the standard or depthwise convolution within the block. The group number of a block refers to the number of groups in the standard or pointwise convolution within the block. When the group number is 2, a channel shuffle \cite{ShuffleNet} operation is applied afterward. The encoder architecture comprises five blocks, generating 25 searchable nodes and creating a search space of approximately $10^{12}$ possible configurations. We employ the same RL-based search algorithm proposed in \cite{MnasNet}. However, in contrast to optimizing for latency, we focus on optimizing MACS since we do not target any specific hardware device.

\begin{figure}
    \centering
    \includegraphics[width=\linewidth]{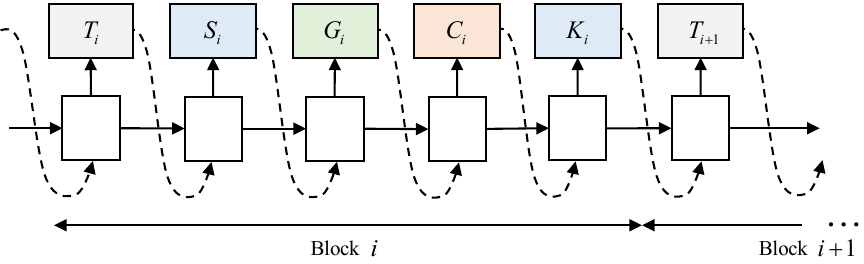}
    \caption{Architecture of the controller used in the reinforcement learning-based search algorithm, with each white block representing a basic cell, composed of an embedding layer followed by an LSTM cell and a sampling layer.}
    \label{fig:controller}
\end{figure}

As shown in Fig.~\ref{fig:nas}, the search process consists of three components: an RNN-based controller, a trainer to obtain the speech quality metric on the validation set, and a multi-objective reward function. The controller serves as the RL agent and contains the same number of cells as nodes, as illustrated in Fig.~\ref{fig:controller}. Each cell is composed of an embedding layer followed by an LSTM cell and a sampling layer. The embedding layer first maps the one-hot action from the previous cell into a high-dimensional space. The LSTM cell then takes this embedded vector as input and generates a logit vector, representing the preferences for each possible action. Finally, the sampling layer uses the logit vector to compute a probability distribution through a softmax function and samples an action based on this distribution. By recursively predicting a sequence of actions, the controller generates a representation that exhibits a one-to-one mapping with the distinct architectures within the search space. 

During the controller training process, a batch of models is first sampled in accordance with the predicted sequence. Subsequently, each sampled model $m$ is trained in parallel on an SE dataset to get its speech quality metric $Q(m)$. The reward value $R(m)$ is computed by taking into account both $Q(m)$ and MACS $M(m)$ via the following formulation:
\begin{equation}
    R(m)=\left(Q(m)-Q_0\right)\times \left( \frac{M(m)}{M_T} \right)^\omega,
    \label{eq:11}
\end{equation}
where $Q_0$ is a baseline score which maps the $Q(m)$ to a reasonable range, $M_T$ denotes the target MACS, and $\omega$ serves as a weight factor for quality-computation trade offs. The $\omega$ is defined as:
\begin{equation}
    \omega =
    \begin{cases}
        \omega_{+}, & \text{if } M(m) > M_T, \\
        \omega_{-}, & \text{otherwise},
    \end{cases}
\end{equation}
where $\omega_{+}$ and $\omega_{-}$ are application-specific constants. Eventually, the parameters of the controller are updated by maximizing the expected reward using proximal policy optimization \cite{PPO}. The aforementioned procedure is repeated until the parameters of the controller converge.

\subsection{Loss function}
We employ a hybrid loss function applied on both the waveform and spectrogram domains to train the SE models, formulated as:
\begin{equation}
  \begin{split}
  \mathcal{L}=\rev{\lambda_1}\mathcal{L}_{\text{SISNR}}(\tilde{s},s) + (1-\rev{\lambda_2}) \mathcal{L}_{\text{mag}}(\tilde{S},S) \\
             + \rev{\lambda_2}\left(\mathcal{L}_{\text{real}}(\tilde{S},S)+\mathcal{L}_{\text{imag}}(\tilde{S},S)\right),
  \end{split}
\end{equation}
where $\tilde{s}$ and $s$ are the enhanced and clean waveforms. $\tilde{S}$ and $S$ are the enhanced and clean spectrograms, respectively. \rev{$\lambda_1$} and \rev{$\lambda_2$} are set to 0.01 and 0.3, respectively, \rev{as these values were empirically determined to balance the contributions of each loss component and were also validated in GTCRN \cite{GTCRN}}. Each term in the aforementioned formula is calculated as follows:
\begin{gather}
  \mathcal{L}_{\text{SISNR}}=-\log_{10}\left( \frac{\left\|s_{t}\right\|^{2}}
  {\left\|\tilde{s}-s_{t}\right\|^{2}} \right);
  s_{t}=\frac{\langle \tilde{s}, s \rangle s}{\left\|s\right\|^{2}},
  \\
  \mathcal{L}_{\text{mag}}(\tilde{S},S)=\mathrm{MSE}(|\tilde{S}|^{0.3},|S|^{0.3}),
  \\
  \mathcal{L}_{\text{real}}(\tilde{S},S)=\mathrm{MSE}(\tilde{S_{r}}/|\tilde{S}|^{0.7},S_{r}/|S|^{0.7}),
  \\
  \mathcal{L}_{\text{imag}}(\tilde{S},S)=\mathrm{MSE}(\tilde{S_{i}}/|\tilde{S}|^{0.7},S_{i}/|S|^{0.7}).
\end{gather}

\section{Experiments}
\subsection{Datasets}

We evaluate our proposed model using two datasets. The first is the large-scale DNS3 dataset \cite{DNS3}, which includes a diverse collection of clean speech, noise, and RIRs. Additionally, we incorporate the Mandarin corpus from DiDiSpeech \cite{DiDiSpeech}. During data mixing, clean speech is convolved with a randomly selected RIR and then mixed with randomly chosen noise clips under an SNR range of -5 to \SI{15}{dB}. The training target is obtained by preserving the first 100 ms of reflections. A total of 720,000 pairs of 10-second noisy-clean data are created for training, while 1,000 pairs are generated for validation and another 1,000 pairs for testing, respectively. For the ablation study, a smaller subset ($10\%$ of the complete dataset, with each utterance cropped to 4 seconds) is sampled to facilitate faster convergence. 

The second dataset is the VCTK-DEMAND dataset \cite{VCTK-DEMAND}, which contains paired clean and pre-mixed noisy speech. The training and test set consists of 11,572 utterances from 28 speakers and 824 utterances from two speakers, respectively. 1,572 utterances in the training set are selected for validation. All the utterances are sampled at \SI{16}{kHz}. 

\subsection{Implementation details}
\subsubsection{Parameter configurations for model prototype}
The BM module adopts the same setup in GTCRN, which maps the 192 high-frequency bands to 64 ERB bands, while keeping the 65 low-frequency bands unaltered, leading to a 129-dimensional compressed feature map. 
\rev{For the ablation study, we construct a model prototype whose encoder comprises five identical blocks, selected from the basic or extended blocks as described in Sec.~\ref{sec:base_blocks}.} All blocks have 16 channels and a kernel size of (3,3). The first two blocks have a stride of 2 in the frequency dimension to downsample the feature map, while the remaining three blocks have a stride of 1. The decoder is a mirror version of the encoder, where all the standard or depthwise convolutional layers are replaced with transposed ones with the same parameter configurations. 

\subsubsection{Training configurations}
The STFT is computed using a Hanning window with a length of 32 ms, a hop length of 16 ms, and an FFT size of 512 points. The log-power spectrogram is fed into the model, which outputs an estimated IRM to obtain the enhanced magnitude. The enhanced speech is then obtained by combining the enhanced magnitude with the noisy phase. The models are trained using the Adam optimizer \cite{Adam} with two different schedulers. For the DNS3 dataset, a linear-warm-up cosine-annealing scheduler is used. The learning rate increases linearly from $10^{-6}$ to $10^{-3}$ over the first 25,000 steps and then decays cosine-wise until 250,000 steps, at which point the training procedure ends. A batch size of 32 is used for the full DNS3 dataset, while a batch size of 8 is used for the DNS3 subset. For the VCTK-DEMAND dataset, a performance-based learning rate scheduler is applied. The learning rate starts at $10^{-3}$ and is halved if the validation loss fails to decrease for 5 consecutive epochs. For this dataset, a batch size of 4 is used.

\subsubsection{Neural architecture search configurations}
The controller cell has an embedding size of 100 and an LSTM hidden size of 200. It is trained using the Adam optimizer with an initial learning rate of 0.001. The multi-processing technique is employed to sample and evaluate multiple architectures concurrently. For each episode, 40 architectures are sampled and trained on 8 RTX 3090 GPUs, with 5 models per GPU and a batch size of 8 for each model. Each model is trained on the DNS3 subset for 20 epochs, using 5,000 randomly selected noisy-clean pairs per epoch. Validation is conducted only during the final three epochs, and the average perceptual evaluation of speech quality (PESQ) \cite{PESQ} score is used as the $Q(m)$ in Eq.~(\ref{eq:11}). The baseline score $Q_0$ is set to 1, and the target MACS $M_T$ is set to \rev{30M}\footnote{\rev{The computational complexity of all models was evaluated using the \texttt{ptflops} toolkit, with a fixed input length of 1 second.}}. The weight constants $\omega_{+}$ and $\omega_{-}$ are set to -0.15 and 0, respectively. If the reward does not increase for 5 consecutive epochs, the learning rate is halved. 

\subsection{\rev{Baselines} and evaluation metrics}
We compare our model with the latest SOTA ultra-lightweight SE models, including GTCRN \cite{GTCRN} and LiSenNet \cite{LiSenNet}. \rev{In addition, we include scaled-down versions of CRN \cite{CRN} (CRN-small) and DPCRN \cite{DPCRN} (DPCRN-small), as well as DeepFilterNet \cite{DeepFilterNet}, to provide a more comprehensive comparison. The CRN-small and DPCRN-small models are implemented by ourselves, following the original papers’ configurations while proportionally reducing the number of channels for model compression. For DeepFilterNet, we use the official public implementation. All baselines are retrained under the same training setup as our UL-UNAS model}.

When comparing on the VCTK-DEMAND dataset, some models from this benchmark are also considered, including FullSubNet \cite{FullSubNet}, FastFullSubNet \cite{FastFullSubNet}, and GaGNet \cite{GaGNet}. We use the statistics reported in the original papers if available; otherwise, we retrain these models using the official implementations with our training configurations to ensure fairness.

We use three intrusive metrics for performance evaluation: the scale-invariant signal-to-noise ratio (SISNR) \cite{SISNR}, the PESQ, and the extended short-time objective intelligibility (ESTOI) \cite{ESTOI}. Additionally, two non-intrusive metrics, DNSMOS P.808 \cite{DNSMOS} and DNSMOS P.835 \cite{DNSMOS-P835}, are employed to provide a more comprehensive comparison. DNSMOS P.835 includes three sub-metrics: OVRL for overall speech quality, SIG for signal quality, and BAK for background noise quality.

\begin{table*}[t]
\centering
\caption{Performance of model prototypes equipped with various basic blocks.}
\renewcommand{\arraystretch}{1.2}
\resizebox{\linewidth}{!}{
\begin{tabular}{c|c|cc|ccc|ccc|c}
\hline\hline
  \multirow{2}{*}{\textbf{IDs}} & 
  \multirow{2}{*}{\textbf{Blocks}} &
  \multirow{2}{*}{\textbf{Params (k)}} & 
  \multirow{2}{*}{\textbf{MACS (M/s)}} & 
  \multirow{2}{*}{\textbf{SISNR}} & 
  \multirow{2}{*}{\textbf{PESQ}} & 
  \multirow{2}{*}{\textbf{ESTOI ($\times100$)}} & 
  \multicolumn{3}{c|}{\textbf{DNSMOS-P.835}} & 
  \multirow{2}{*}{\textbf{DNSMOS-P.808}} \\ \cline{8-10}
  & 
  &  
  & 
  &   
  &   
  &       
  & \textbf{OVRL} & \textbf{SIG} & \textbf{BAK}
  &      \\ \hline
1 & Conv $\times$ 5  & 52.12  & 57.96  & \textbf{11.12} & \textbf{2.05}  & \textbf{74.83} & \textbf{2.64} & \textbf{2.98} & \textbf{3.84} & \textbf{3.50} \\ \hline
2 & DWS $\times$ 5   & \textbf{37.23}  & \textbf{23.72}  & 10.80 & 1.99  & 74.08 & 2.59 & 2.94 & 3.81 & 3.46 \\
3 & Ghost $\times$ 5 & 43.39  & 37.82  & 10.98 & 2.02  & 74.49 & 2.60 & 2.95 & 3.81 & 3.46 \\
4 & Rep $\times$ 5   & 37.23  & 23.72  & 10.87 & 2.00  & 74.29 & 2.59 & 2.94 & 3.80 & 3.45 \\ \hline
5 & MB $\times$ 5    & 39.98  & 30.68  & 10.86 & 2.03  & 74.19 & 2.61 & 2.95 & 3.82 & 3.47 \\
6 & Star $\times$ 5  & 39.97  & 31.46  & 10.73 & 1.98  & 74.05 & 2.57 & 2.92 & 3.79 & 3.44  \\ \hline\hline
\end{tabular}
}
\label{tab:1}
\end{table*}

\begin{table*}[t!]
\centering
\caption{
\rev{Ablation study results of the APReLU activation and comparison with other activation functions. The values in parentheses under the ``Act.'' and ``Affine'' columns specify the dimensionality of the corresponding parameters.}
}
\renewcommand{\arraystretch}{1.2}
\resizebox{\linewidth}{!}{
\begin{tabular}{c|cc|cc|ccc|ccc|c}
\hline\hline
  \multirow{2}{*}{\textbf{IDs}} & 
  \multirow{2}{*}{\rev{\textbf{Act.}}} & 
  \multirow{2}{*}{\textbf{Affine}} & 
  \multirow{2}{*}{\textbf{Params (k)}} & 
  \multirow{2}{*}{\textbf{MACS (M/s)}} & 
  \multirow{2}{*}{\textbf{SISNR}} & 
  \multirow{2}{*}{\textbf{PESQ}} & 
  \multirow{2}{*}{\textbf{ESTOI ($\times100$)}} & 
  \multicolumn{3}{c|}{\textbf{DNSMOS-P.835}} & 
  \multirow{2}{*}{\textbf{DNSMOS-P.808}} \\ \cline{9-11}
   &  
   &
   &
   &
   &
   &
   &
   & \textbf{OVRL} & \textbf{SIG} & \textbf{BAK} &
   \\ \hline
1  & \rev{PReLU (1)}  & \rev{\ding{55}}  & 37.23  & 23.72 & 10.80 & 1.99 & 74.08 & 2.59 & 2.94 & 3.81  & 3.46 \\
2  & \rev{PReLU (C)}  & \rev{\ding{55}}  & 37.50  & 23.72 & 10.91 & 2.01 & 74.47 & 2.58 & 2.94 & 3.78  & 3.45  \\
3  & \rev{PReLU (CxF)}  & \rev{\ding{55}}  & 49.86  & 23.72 & 10.97 & 2.02  & 74.41 & 2.60 & 2.95 & 3.81 & 3.48  \\ \hline
4  & \rev{PReLU (1)}  & \rev{\ding{51} (1)}  & 37.27 & 23.72  & 10.75 & 1.99 & 74.04 & 2.60 & 2.94 & 3.81 & 3.47  \\
5  & \rev{PReLU (C)} & \rev{\ding{51} (C)}  & 38.08  & 23.72 & 10.67 & 2.00 & 73.96 & 2.60 & 2.95 & 3.81 & 3.47  \\
6  & \rev{PReLU (C)}  & \rev{\ding{51} (CxF)}  & 62.79  & 23.72 & 10.97 & 2.04 & 74.50 & \textbf{2.63} & \textbf{2.97} & \textbf{3.84}  & \textbf{3.50} \\
7  & \rev{PReLU (CxF)}  & \rev{\ding{51} (CxF)} & 75.14 & 23.72 & \textbf{11.01} & \textbf{2.05}  & \textbf{74.57}  & 2.61 & 2.95 & 3.81 & 3.50  \\ \hline
8  & \rev{GELU} & \rev{\ding{55}} & \textbf{\rev{37.22}}  & \rev{23.72}  & \rev{10.87}  &  \rev{1.99}  & \rev{74.02}  & \rev{2.60}  & \rev{2.94} &  \rev{3.82}  &  \rev{3.47} \\
9  & \rev{GELU} & \rev{\ding{51} (CxF)} & \rev{62.50}  & \rev{23.72}  & \rev{10.94}  &  \rev{2.03}  & \rev{74.35}  & \rev{2.62}  & \rev{2.96} &  \rev{3.83}  &  \rev{3.50} \\
10  & \rev{Swish} & \rev{\ding{55}} & \rev{37.50}  & \textbf{\rev{22.92}}  & \rev{10.76}  &  \rev{1.99}  & \rev{74.05}  & \rev{2.60}  & \rev{2.94} &  \rev{3.82}  &  \rev{3.47} \\
11  & \rev{Swish} & \rev{\ding{51} (CxF)} & \rev{62.79}  & \textbf{\rev{22.92}}  & \rev{10.97}  &  \rev{2.03}  & \rev{74.35}  & \rev{2.62}  & \rev{2.96} &  \rev{3.82}  &  \rev{3.50} \\

\hline\hline
\end{tabular}
}
\label{tab:2}
\end{table*}

\begin{table*}[t]
\centering
\caption{Ablation study results of the cTFA module.}
\renewcommand{\arraystretch}{1.2}
\resizebox{\linewidth}{!}{
\begin{tabular}{c|cc|cc|ccc|ccc|c}
\hline\hline
  \multirow{2}{*}{\textbf{IDs}} & 
  \multirow{2}{*}{\textbf{\rev{TA}}} & 
  \multirow{2}{*}{\textbf{\rev{FA}}} & 
  \multirow{2}{*}{\textbf{Params (k)}} & 
  \multirow{2}{*}{\textbf{MACS (M/s)}} & 
  \multirow{2}{*}{\textbf{SISNR}} & 
  \multirow{2}{*}{\textbf{PESQ}} & 
  \multirow{2}{*}{\textbf{ESTOI ($\times100$)}} & 
  \multicolumn{3}{c|}{\textbf{DNSMOS-P.835}} & 
  \multirow{2}{*}{\textbf{DNSMOS-P.808}} \\ \cline{9-11}
   &  
   &
   &
   &
   &
   &
   &
   & \textbf{OVRL} & \textbf{SIG} & \textbf{BAK} &
   \\ \hline
1  & \ding{55} & \ding{55} & \textbf{37.23} & 23.72  & 10.80 & 1.99 & 74.08 & 2.59 & 2.94 & 3.81 & 3.46  \\
2 & \ding{51} & \ding{55} & 85.22  & 26.87  & 11.38 & 2.09 & 75.44 & 2.63 & 2.97 & 3.83 & 3.50  \\
3 & \ding{55} & \ding{51} & \rev{39.99}  & \rev{26.43}  & \rev{10.85} & \rev{2.00} & \rev{74.12} & \rev{2.61} & \rev{2.95} & 3.82 & \rev{3.48}  \\
4 & \ding{51} & \ding{51} & \rev{87.98} & \textbf{\rev{29.59}}  & \textbf{\rev{11.66}} & \textbf{\rev{2.15}}  & \textbf{\rev{76.20}} & \textbf{\rev{2.65}} & \textbf{\rev{2.99}} & \textbf{\rev{3.85}} & \textbf{\rev{3.53}}  \\ \hline\hline
\end{tabular}
}
\label{tab:3}
\end{table*}

\begin{figure}
    \centering
    \includegraphics[width=0.95\linewidth]{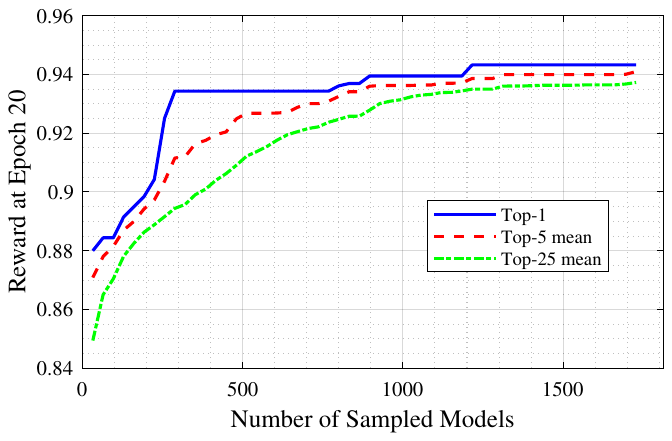}
    \caption{Rewards of top models during the process of NAS.}
    \label{fig:reward}
\end{figure}

\begin{table}[t]
\tiny
\centering
\caption{Optimal network configuration discovered by NAS.}
\renewcommand{\arraystretch}{1.2}
\resizebox{0.9\linewidth}{!}{
\begin{tabular}{l|l}
\hline\hline
  Types    & {[}XConv, XMB, XDWS, XMB, XDWS{]}       \\ \hline
  Strides  & {[}2, 2, 1, 1, 1{]}                     \\ \hline
  Groups   & {[}1, 2, 2, 2, 2{]}                     \\ \hline
  Channels & {[}12, 24, 24, 32, 16{]}                \\ \hline
  Kernel  & {[}(3,3), (2,3), (2,3), (1,5), (1,5){]} \\ \hline\hline
\end{tabular}
}
\label{tab:4}
\end{table}

\begin{table*}[t]
\centering
\caption{Performance of models equipped with diverse combinations of blocks.}
\renewcommand{\arraystretch}{1.2}
\resizebox{\linewidth}{!}{
\begin{tabular}{c|c|cc|ccc|ccc|c}
\hline\hline
  \multirow{2}{*}{\textbf{IDs}} & 
  \multirow{2}{*}{\textbf{Blocks}} &
  \multirow{2}{*}{\textbf{Params (k)}} & 
  \multirow{2}{*}{\textbf{MACS (M/s)}} & 
  \multirow{2}{*}{\textbf{SISNR}} & 
  \multirow{2}{*}{\textbf{PESQ}} & 
  \multirow{2}{*}{\textbf{ESTOI ($\times100$)}} & 
  \multicolumn{3}{c|}{\textbf{DNSMOS-P.835}} & 
  \multirow{2}{*}{\textbf{DNSMOS-P.808}} \\ \cline{8-10}
  & 
  &  
  & 
  &   
  &   
  &       
  & \textbf{OVRL} & \textbf{SIG} & \textbf{BAK}
  &      \\ \hline
1  & XConv $\times$ 5  & \rev{114.56}  & \rev{63.11}  & \rev{11.96}  & \rev{2.22} & \rev{77.08} & \textbf{\rev{2.69}} & \textbf{\rev{3.02}} & \textbf{\rev{3.86}} & \rev{3.53} \\ 
2  & XDWS $\times$ 5   & \textbf{\rev{113.53}}  & \textbf{\rev{28.00}}  & \rev{11.69}  & \rev{2.16} & \rev{76.47} & \rev{2.66} & \rev{2.99} & \rev{3.84} & \rev{3.53} \\ 
3  & XMB $\times$ 5    & \rev{116.53}  & \rev{34.94}  & \rev{11.85}  & \rev{2.20} & \rev{77.01} & \rev{2.67} & \rev{3.00} & \rev{3.84} & \rev{3.53} \\  \hline
4  & NAS-discovered         & \rev{171.33}  & \rev{34.91}  & \textbf{\rev{12.07}}  & \textbf{\rev{2.25}} & \textbf{\rev{77.69}} & \textbf{\rev{2.69}} & \rev{3.01} & \textbf{\rev{3.86}} & \textbf{\rev{3.55}} \\ \hline\hline

\end{tabular}
}
\label{tab:5}
\end{table*}

\begin{table*}[tbp]
\centering
\caption{Performance comparison on the DNS3 test set. LA represents look ahead. \rev{All models are retrained with the same training configuration as UL-UNAS.}}
\renewcommand{\arraystretch}{1.2}
\resizebox{\linewidth}{!}{
\begin{tabular}{c|c|cc|ccc|ccc|c}
\hline\hline
  \multirow{2}{*}{\textbf{Models}} &
  \multirow{2}{*}{\textbf{LA (ms)}} & 
  \multirow{2}{*}{\textbf{Params (k)}} & 
  \multirow{2}{*}{\textbf{MACS (M/s)}} & 
  \multirow{2}{*}{\textbf{SISNR}} & 
  \multirow{2}{*}{\textbf{PESQ}} & 
  \multirow{2}{*}{\textbf{ESTOI ($\times 100$)}} &
  \multicolumn{3}{c|}{\textbf{DNSMOS-P.835}} & 
  \multirow{2}{*}{\textbf{DNSMOS-P.808}} \\ \cline{8-10}
  &                                 
  &                    
  &                
  &              
  &          
  & 
  & \textbf{OVRL} & \textbf{SIG} & \textbf{BAK}
  &
  \\ \hline
Noisy  & -  & -  & -  & 5.61  & 1.40  & 66.90 & 1.63 & 2.05 & 1.86 & 2.82  \\
\rev{CRN-small} & \rev{0} & \rev{195} & \rev{35} & \rev{11.34} & \rev{2.07} & \rev{75.22} & \rev{2.62} & \rev{2.95} & \rev{3.84} & \rev{3.49} \\ 
\rev{DPCRN-small} & \rev{0} & \rev{\textbf{15}} & \rev{38} & \rev{10.67} & \rev{2.01} & \rev{74.29} & \rev{2.59} & \rev{2.94} & \rev{3.84} & \rev{3.49} \\ 
\rev{DeepFilterNet} & \rev{0} & \rev{1780} & \rev{358} & \rev{12.20} & \rev{2.13} & \rev{75.92} & \rev{\textbf{2.73}} & \rev{\textbf{3.07}} & \rev{3.86} & \rev{3.43} \\
GTCRN  & 0  & 48 & \textbf{34} & 11.44 & 2.10 & 75.71 & 2.63 & 2.97 & 3.82  & 3.50 \\
LiSenNet & 32 & 37 & 56 & 12.12 & 2.23 & 77.02 & 2.70 & 3.04 & 3.85 & 3.55 \\ \hline
UL-UNAS (proposed)    & 0  & \rev{171} & \textbf{\rev{35}}  & \textbf{\rev{12.30}} & \textbf{\rev{2.30}} & \textbf{\rev{78.24}} & \rev{2.71} & \rev{3.05} & \textbf{\rev{3.88}}  & \textbf{\rev{3.56}} \\ \hline\hline
\end{tabular}
}
\label{tab:6}
\end{table*}

\begin{table}[t]
\centering
\caption{Performance comparison on the VCTK-DEMAND test set. The values in \textbf{BOLD} indicate the best results in each metric.}
\renewcommand{\arraystretch}{1.2}
\resizebox{\linewidth}{!}{
\begin{tabular}{c|cc|ccc}
\hline\hline
  \textbf{Models} &
  \textbf{Params (M)} &
  \textbf{MACS (G/s)} &
  \textbf{SISNR} &
  \textbf{PESQ} & 
  \textbf{STOI} \\ \hline
Noisy & - & - & 8.45 & 1.97 & 0.921 \\
RNNoise (2018) & 0.06 & 0.04 & - & 2.29 & - \\
PercepNet (2020) & 8.00 & 0.80 & - & 2.73 & - \\
FullSubNet\textsuperscript{$\ast$} (2021) & 5.64 & 30.87 & - & 2.89 & 0.940 \\
DeepFilterNet (2022) & 1.80 & 0.35 & 16.63 & 2.81 & \textbf{0.942} \\
GaGNet (2022) & 5.95 & 1.65 & - & 2.94 & 0.940  \\
Fast FullSubNet\textsuperscript{$\ast$} (2023) & 6.84 & 4.14 & - & 2.79 & 0.935 \\
GTCRN (2024) & 0.05 & \textbf{0.03} & \textbf{18.83} & 2.87 & 0.940      \\
LiSenNet\textsuperscript{$\dagger$} (2024) & \textbf{0.04} & 0.06 & - & 2.95 & 0.937 \\
LiSenNet\textsuperscript{$\ddagger$} (2024) & \textbf{0.04} & 0.06 & - & 3.07 & 0.939 \\ \hline
UL-UNAS\textsuperscript{$\dagger$} (proposed) & 0.17 & \textbf{0.03} & 18.25 & 2.96 & 0.941 \\ 
UL-UNAS\textsuperscript{$\ddagger$} (proposed) & 0.17 & \textbf{0.03} & 18.48 & \textbf{3.09} & 0.941 \\ 
\hline\hline
\end{tabular}
}
\begin{flushleft}
\textsuperscript{$\ast$} Metrics are porvided by \cite{LiSenNet}. \\
\textsuperscript{$\dagger$} The model is trained without PESQ loss. \\
\textsuperscript{$\ddagger$} The model is trained with additional PESQ loss. \\
\end{flushleft}
\label{tab:7}
\end{table}

\section{Results and analysis}
\subsection{Ablation study}
\subsubsection{Effect of \rev{base} blocks}
\label{sec:effects_base_blocks}
To explore the potential of various convolutional blocks, we compare the performance of different model prototypes, each equipped with distinct convolutional blocks. The results are presented in Table \ref{tab:1}. It can be observed that when the encoder uses only Conv blocks, it achieves fairly good performance, but at the cost of a significantly heavier computational load, as is shown in ID1. 
On the other hand, using efficient convolutional blocks provides a trade-off between computation and performance. Among the blocks with two convolutional layers (from ID2 to ID4), the DWS block stands out for its significant reduction in computational cost, albeit with an inevitable decline in performance. In contrast, the Ghost block achieves a marginal performance advantage but at the expense of the highest computational cost. Meanwhile, the Rep block, which employs a reparameterization technique, fails to deliver meaningful performance improvements despite its additional operational complexity. Given these observations, we conclude that the DWS block offers a more efficient balance between performance and computational cost. In the case of blocks with three convolutional layers (ID5 and ID6), the MB block demonstrates a clear performance advantage with fewer MACS. Based on these results, the Conv block, DWS block, and MB block are selected as representatives for subsequent experiments.

\subsubsection{Effect of affine PReLU}
In this section, we investigate the application of APReLU to the DWS block to demonstrate its effectiveness. Specifically, we design the following experiments: (1) analyzing the impact of the slope-controlling parameter shape in PReLU, and (2) examining the effect of the shapes of the two affine parameters. The results are presented in Table~\ref{tab:2}. Note that the computational overhead of affine transformations is disregarded, as it is minimal and negligible. 

It can be observed that extending the shape of the slope-controlling parameter yields slight performance improvements from ID1 to ID3, likely due to more fine-grained activations across different dimensions. Similarly, the performance boost from the affine transformation is also closely tied to the parameter shapes. When the affine parameters are frequency-independent, the gains are negligible, as shown in ID4 and ID5. However, when the parameters are frequency-dependent, more \rev{noticeable} performance gains are observed, as shown in ID6 and ID7\rev{, highlighting the importance of frequency-dependent affine recalibration}. Comparing ID6 and ID7, the improvements in intrusive metrics are minimal, and the DNSMOS scores even degrade. This suggests that using both a frequency-dependent slope-controlling parameter and affine parameters appears somewhat redundant. Overall, the APReLU activation provides \rev{consistent and meaningful} performance gains, with a 0.05 improvement in PESQ and a 0.04 increase in OVRL and DNSMOS-P.808.

\rev{To enable a comprehensive comparison, we also incorporate existing advanced activation functions, including GELU \cite{GELU} and Swish \cite{Swish}. As shown by ID8 and ID10 versus ID6, APReLU consistently achieves superior performance. Furthermore, applying our affine transformation to GELU (ID9) and Swish (ID11) also yields measurable gains (PESQ: +0.04, DNSMOS OVRL: +0.02). This result confirms that the frequency-dependent affine recalibration is a general-purpose design principle, consistent with its effectiveness on PReLU.}

\subsubsection{Effect of causal time-frequency attention}
In this section, we explore the application of cTFA to the DWS block to demonstrate its effectiveness. Specifically, we design the following experiments: (1) employing only the \rev{TA} module, (2) employing only the \rev{FA} module, and (3) employing both the \rev{TA} and \rev{FA} modules. The results are presented in Table~\ref{tab:3}, which shows that the TA module significantly enhances the model's performance, achieving a 0.1 increase in PESQ, as observed in ID1 and ID2. In contrast, the exclusive application of the FA module provides little improvement, as shown in ID3. However, by integrating both attention branches, the model achieves the best performance, with a total PESQ improvement of \rev{0.16}, as demonstrated in ID4. This suggests that the frequency-dependent characteristics of the \rev{FA} module complement the \rev{TA} module. Notably, while both the \rev{TA} and \rev{FA} modules introduce additional parameters, the resulting computational overhead remains modest and acceptable.

\subsubsection{Effect of neural architecture search}
In this section, we first explore the application of both APReLU and cTFA to the three basic blocks, forming extended blocks that serve as candidate blocks for NAS. We present the trend of reward changes with respect to the number of sampled models in Fig.~\ref{fig:reward}. It can be observed that the reward of the top-1 model improves steadily as the number of sampled models increases. After sampling more than 1200 models, the rewards of the top-1, top-5, and top-25 models reach close proximity, indicating that the NAS process is nearing convergence. The entire search lasts for \rev{only} approximately 2 \rev{GPU} days, \rev{underscoring the practicality of our approach for rapid, automated ultra-lightweight model design.} The optimal network configuration discovered by NAS is presented in Table~\ref{tab:4}. The performance results of the discovered model, along with models equipped with the three extended blocks, are shown in Table~\ref{tab:5}. By comparing ID1, ID2, and ID5 in Table~\ref{tab:1} with ID1, ID2, and ID3 in Table~\ref{tab:5}, we observe that the application of both APReLU and cTFA leads to substantial improvements across all metrics, highlighting the broad effectiveness of both modules. Moreover, \rrev{applying} the NAS technique \rrev{(ID4)} further enhances performance compared to the overall best manually designed prototype network (ID3)\rrev{, as evidenced by a 0.05 gain in PESQ and a 0.02 increase in DNSMOS OVRL. Note that the improvements remain relatively stable across different random seeds, highlighting the robustness and superiority of the NAS-discovered architecture}. Interestingly, \rrev{given strict MACS constraints}, NAS tends to favor models with a larger number of parameters, even though we impose no restrictions on parameter count. This somewhat suggests the importance of having more parameters in improving model performance. Notably, due to a reduced total number of \rev{convolutional} layers, the NAS-discovered architecture also results in lower inference latency \rev{compared with the XMB$\times$5-based version}.

\subsection{Comparison with baselines}
We train UL-UNAS and other baseline models on the full DNS3 dataset (2,000 hours), and evaluate them on our synthesized test set. Results are shown in Table \ref{tab:6}. \rev{Compared with typical U-Net models such as CRN-small and DPCRN-small, UL-UNAS yields consistent and substantial improvements across all metrics while maintaining a similar computational budget. Notably, despite requiring nearly the same MACS as GTCRN (35M vs. 34M), UL-UNAS delivers clear advantages, achieving a remarkable PESQ gain of 0.20, an OVRL increase of 0.08, and a DNSMOS-P.808 improvement of 0.06.
UL-UNAS also achieves competitive or superior performance compared with DeepFilterNet, even though the latter is more than 10$\times$ larger in MACS and over 10$\times$ larger in parameters. 
Furthermore, UL-UNAS reaches or surpasses LiSenNet while avoiding its additional 32 ms look-ahead introduced by the dual-iteration Griffin–Lim phase refinement. Overall, UL-UNAS establishes a new SOTA trade-off among lightweight, causal SE models, offering the strongest performance with strictly real-time, zero-look-ahead processing.}

For a more comprehensive comparison with other models, we also train UL-UNAS on the VCTK-DEMAND dataset, with the results presented in Table~\ref{tab:7}. Notably, we provide two sets of performance results for UL-UNAS: one from the model trained with the standard loss function and the other from the model trained with an additional PESQ loss\footnote{The PESQ loss is implemented by \texttt{torch-pesq} packet: \url{https://github.com/audiolabs/torch-pesq}.}. This approach ensures alignment with the results reported in the original LiSenNet paper and facilitates a fair comparison. It is evident that UL-UNAS outperforms LiSenNet in both cases. Furthermore, UL-UNAS demonstrates the best performance among all the models while maintaining lower computational costs and a relatively compact model size.

\section{Conclusions}
\label{sec:conclusion}
In this paper, we present UL-UNAS, an ultra-lightweight SE model that requires only \rev{35M} MACS. By systematically exploring efficient convolutional blocks and introducing innovative components, including the APReLU and cTFA modules, we construct a more powerful U-Net family that excels in computational efficiency and enhancement quality. Furthermore, the employment of NAS enables the discovery of an optimized architecture tailored to real-time constraints \rev{in a short search time, presenting a practical and scalable approach for automating lightweight model design.} Through these contributions, UL-UNAS not only surpasses SOTA ultra-lightweight models but also demonstrates competitive results against more resource-intensive baseline models, making it a robust solution for real-time SE applications.

\bibliographystyle{IEEEtran}
\bibliography{ref.bib}

\end{document}